\documentclass[aps]{revtex4}
\usepackage{graphicx}
\def\vec#1{\mbox{\scriptsize\bf #1}}         
\def\p#1{\phantom{#1}}
\begin{document}
\title{New solution for the polaron problem}

\author{I.D.Feranchuk\footnote{Corresponding
author, fax:+375 (17) 220 23 53, e-mail: fer@open.by} and \fbox{L.I.Komarov}}

\address{Belarusian State University, Nezavisimosti Av., 4, 220080
Minsk, Republic of Belarus
}


\begin{abstract}

New variational ansatz for the large-radius Fr\"ohlich polaron
is considered. The corresponding operator estimation for the
energy of polaron proves to
be very similar to the result found by Feynman on the basis of
the variational principle for the functional integral of the system.
It allows us to
make clear the problem of ``phase transition'' and to anylize
the structure of the localized state of the polaron in the
intermediate coupling regime. It is shown that the analogous state
exists also for the model of particle-field interaction with the
divergent perturbation theory.
\end{abstract}

\maketitle
\bigskip

{\em Classification codes}: PACS  71.38.+i; 63.20.Kr


\newpage
\section{Introduction}

It is well known at present that the polaron problem has much
more broad significance than simply a model for the partial form
of interaction
between electron and phonons in the ionic crystal as it was
introduced by Fr\"ohlich \cite{1}. Hamiltonian of the polaron
is considered mostly as a fundamental model for interaction between
particle and quantum field where various non-perturbative methods
for the quantum field theory can be verified in the entire range
of the coupling constant $\alpha$.

Like any other quantum system the polaron can be described
both on the basis of the solution of the Schr\"odinger equation
and by means of calculation of the corresponding Feynman functional
integral. The first way allowed to introduce the idea of
the self-localized polaron \cite{2} and to find the exact
asymptotic value for the ground state energy $E_{0}$ of the system
in the limit $\alpha \rightarrow \infty$ \cite{3}. But
the great advantage of the Feynman variational principle for the
path integrals is the possibility to calculate $E_{0}(\alpha)$
as the continuous function for any $\alpha$ and to find the
lowest estimation for the polaron binding energy in the
intermediate coupling regime \cite{4}.

There were a lot of attempts to calculate the ground state energy
by means of
variational principle for the Schr\"odinger equation (usually it is
called the operator calculation) and some of the
trial functions led to peculiarity for the function $E_{0}(\alpha)$
near the point $\alpha \simeq 7$. These results caused the discussion
about existance of the ``phase transition'' between two qualitatively
different states of the polaron (see review \cite{5} on this problem).
In the series of papers cited in \cite{5} it was proved that
the function $E_{0}(\alpha)$ was analytical for any $\alpha$
and ``phase transition'' did not exist. But it is important to stress
that this strict proof did not consider the constructive
computational algorithm for analysis of the self-localized
and another states
of the polaron.

In the present paper we found new variational ansatz for the
ground state vector for the considered system which leads to
the function $E_{0}(\alpha)$ fairly well coincided with
Feynman's calculation.
It seems to us that the results make the
question about the ground states of the polaron much more
clear and descriptive.

It will be shown below that there are two collective
states of the polaron system which describe the ``dressed'' electron
and do not connect with excitation of the free phonons.
In general case each of them is the mixture of delocalized and
self-localized states of the ``bare'' electron.
The energies of
these states are splitted so that the both
eigenvalues prove to be the continious functions of
$\alpha$ and the ground state energy is similar to the result
found by Feynman. In the limit $\alpha \rightarrow \infty$
the corresponding wave function
 transforms eventually to the self-localized state.
But it is important that both states exist in the entire
range of $\alpha$ and a pseudo-intersection between these
energy levels could be the reason for imaginary ``phase transition''
for some sort of variational estimations.

Our analysis also showed that the conception of self-localized state
proved to be applicable for more general models of the particle-field
interaction even in the case when usual perturbation theory includes
both the infrared and ultraviolet divergencies.

\section{Description of two  basic states}

We use the standard form of the Fr\"ohlich Hamiltonian
($ \hbar = m = \omega = 1$)

\begin{equation}
\label{eq1}
\hat H = -\frac{1}{2}\Delta + \sum_{\vec k} a^{+}_{\vec k}
a^{\p+}_{\vec k} + 2^{3/4}\left(\frac{\pi\alpha}{\Omega}\right)^{1/2}
\sum_{\vec k}\frac{1}{k}\Bigl(e^{i\vec k \vec r} a^{\p+}_{\vec k} +
e^{-i\vec k \vec r} a^{+}_{\vec k}\Bigr).
\end{equation}
Here $\bf r$ is the coordinate of electron; $ a^{\p+}_{\vec k}$ and
$a^{+}_{\vec k}$ are the operators of annihilation and creation
of the phonons with the momentum $\bf k$; $\alpha$ is the
dimensionless coupling constant.

The problem is to find the common solution of the Schr\"odinger equation

\begin{equation}
\label{eq2}
\hat H \vert\Psi\rangle  = E(\alpha, {\bf P}) \vert\Psi\rangle,
\end{equation}

and the equation corresponding to the conservation law of the total
momentum

\begin{equation}
\label{eq3}
{\bf \hat P} \vert\Psi\rangle = \Bigl(-i \nabla +  \sum_{\vec k} {\bf k}
a^{+}_{\vec k} a^{\p+}_{\vec k}\Bigr) \vert\Psi\rangle  = {\bf P} \vert\Psi\rangle.
\end{equation}

Let us consider now two well known approximate variational solutions
of these equations which we will use as a basis for our ansatz. We
remind shortly the algorithms for obtaining mentioned solutions.

The first one is defined as follows \cite{6}
\begin{equation}
\label{eq4}
\vert\Psi^{(D)}_{\vec P}({\bf r})\rangle=\frac{1}{\sqrt{\Omega}}
\exp\left(i{\bf Pr}+\sum_{\vec k}v^{\p+}_{\vec k}({\bf P})
\Bigl(e^{-i\vec k\vec r}a^+_{\vec k}-
e^{i\vec k\vec r}a^{\p+}_{\vec k}\Bigr)\right)\vert 0\rangle;\quad
a^{\p+}_{\vec k}\vert 0\rangle=0.
\end{equation}

We call this state delocalized (D) because its structure
corresponds to the polaron where the cloud of the virtual phonons
follows the freely moving electron. The state (\ref{eq4}) is
the eigenfunction for the operator of total momentum and
the parameters $v_{\vec k}({\bf P})$  are defined by the minimum
condition of the energy found as the average of the Hamiltonian
(\ref{eq1}) with respect to this state \cite{7}

\begin{equation}
\label{eq5}
v^{\p+}_{\vec k}({\bf P})=-2^{7/4}\left(\frac{\pi\alpha}{\Omega}\right)^{1/2}
\frac{1}{k[k^2-2({\bf P-Q}){\bf k}+2]},\quad {\bf Q}=\sum_{\vec k}{\bf k}
v^2_{\bf k}({\bf P}).
\end{equation}







For simplicity we'll consider all further calculations for the
ground state with ${\bf P} = 0$ when the energy is defined by well
known simple formula

\begin{equation}
\label{eq8}
E_{D}=H_{DD}=\int d{\bf r}\langle\Psi^{(D)}_{\vec 0}({\bf r})\vert\hat H\vert
\Psi^{(D)}_{\vec 0}({\bf r})\rangle=-\alpha.
\end{equation}


The second basic state is generated by the product ansatz which
was firstly introduced by Pekar \cite{2},
who realized Landau`s \cite{8} suggestion that an electron in an ionic crystal
can be trapped by the static lattice distorsion produced by the electron
itself.
This product ansatz was proved to be the
asymptotically exact solution of the Schr\"odinger equation in the
limit $\alpha \rightarrow \infty$ as it was shown by Bogoliubov \cite{3}.

We can write it as follows \cite{3}

\begin{equation}
\label{eq9}
\vert \Psi({\bf r},{\bf R}) \rangle = \phi({\bf r-R}) \exp
\left(\sum_{\vec k}u^{\p+}_{\vec k}
\Bigl(e^{-i\vec k\vec R}a^{+}_{\vec k} -
e^{i\vec k\vec R}a^{\p+}_{\vec k}\Bigr)\right)
\vert 0\rangle ;\quad
\int d{\bf r}\vert\phi({\bf r})\vert^2=1.
\end{equation}
Here {\bf R} is the arbitrary point in the space where
the electron is localized.
The classical component of the phonon field $u_{\vec k}$
is also defined by the variation of the energy functional and
has the following form (\cite{3})

\begin{equation}
\label{eq10}
u_{\vec k} = -2^{3/4}\left(\frac{\pi\alpha}{\Omega}\right)^{1/2}
\frac{1}{k} \int d{\bf r} \vert \phi({\bf r}) \vert ^2
e^{-i \vec k \vec r},
\end{equation}
and the wave function of the electron in the ground state
should be found as the solution
of the following nonlinear equation
with eigenvalue $E_P$

\begin{equation}
\label{eq11}
[ - \frac{1}{2} \Delta - 2^{1/2}\alpha \int d{\bf r}_{1}
\frac{\vert \phi({\bf r}_{1}) \vert ^2}{\vert {\bf r}_{1} -
{\bf r} \vert} + 2^{-1/2}\alpha\int d{\bf r}_1\int d{\bf r}_2
\frac{\vert \phi({\bf r}_1)\vert^2\vert\phi({\bf r}_2)\vert^2}{\vert{\bf r_1-r_2}\vert}
- E_P ]  \phi({\bf r}) = 0.
\end{equation}

In spite of the exact solution of this equation was found numerically
\cite{9} we will use for further calculation the Pekar`s function
$\phi_0({\bf r})$ which is a very good approximation \cite{9,7}

\begin{equation}
\label{eq12}
\phi_0({\bf r}) = \sqrt{\frac{8}{14+42c+45c^2}}(1+br+cb^2r^2)e^{-br},
\end{equation}
with
$$
b=0.931307\alpha,\quad c=0.451668.
$$
The optimal value for the variational parameter $b, c$ leads to the
polaron binding energy
$$
E_P=-0.108504\alpha^2
$$
which is very closed to the accurate value \cite{9,7} $-0.108513\alpha^2$.



It is clear that the set of the state vectors (\ref{eq9}) correspond to the
localized states
(LS) of the electron
which can be considered as the states with local breakdown of
the translational symmetry of the system in the sense that
was discussed in the fundamental Bogoliubov`s paper \cite{10}.
The energy of the system should not depend on the point of the
electron localization ${\bf R}$ because of this symmetry, and
in order to
reconstruct it let us use \cite{11,12} the linear combination
of degenerate states $\vert\Psi({\bf r},{\bf R})\rangle$


\begin{equation}
\label{eq13}
\vert\Psi^{(L)}_{\vec P}({\bf r})\rangle=\frac{1}{N_{\vec P}\sqrt{\Omega}}
\int d{\bf R}e^{i\vec P\vec R}\vert\Psi({\bf r},{\bf R})\rangle,
\end{equation}
with
$$
N^2_{\vec P}=\int d{\bf R}\int d{\bf r} \phi({\bf r})\,\phi({\bf r+R})
\exp[\sum_{\vec k}u^2_{\vec k}(e^{-i\vec k\vec R}-1)].
$$


This state vector
leads to the corresponding estimation
for the polaron binding energy (we fulfilled real caculations
for the case ${\bf P} = 0$ and used (\ref{eq12}))

\begin{equation}
\label{eq14}
E_{L} = H_{LL}=\int d{\bf r}\,\langle \Psi^{(L)}_{\vec 0}({\bf r}) \vert
 \hat H \vert \Psi^{(L)}_{\vec 0}({\bf r})\rangle
\end{equation}

Fig.1 compares the functions $E_{D}(\alpha)$, $E_{L}(\alpha)$ with the
value $E_P(\alpha)$ calculated with the state vector (\ref{eq9})
without taking into account the translational symmetry.

\begin{figure}[ht]
\includegraphics[width=11cm,height=15cm]{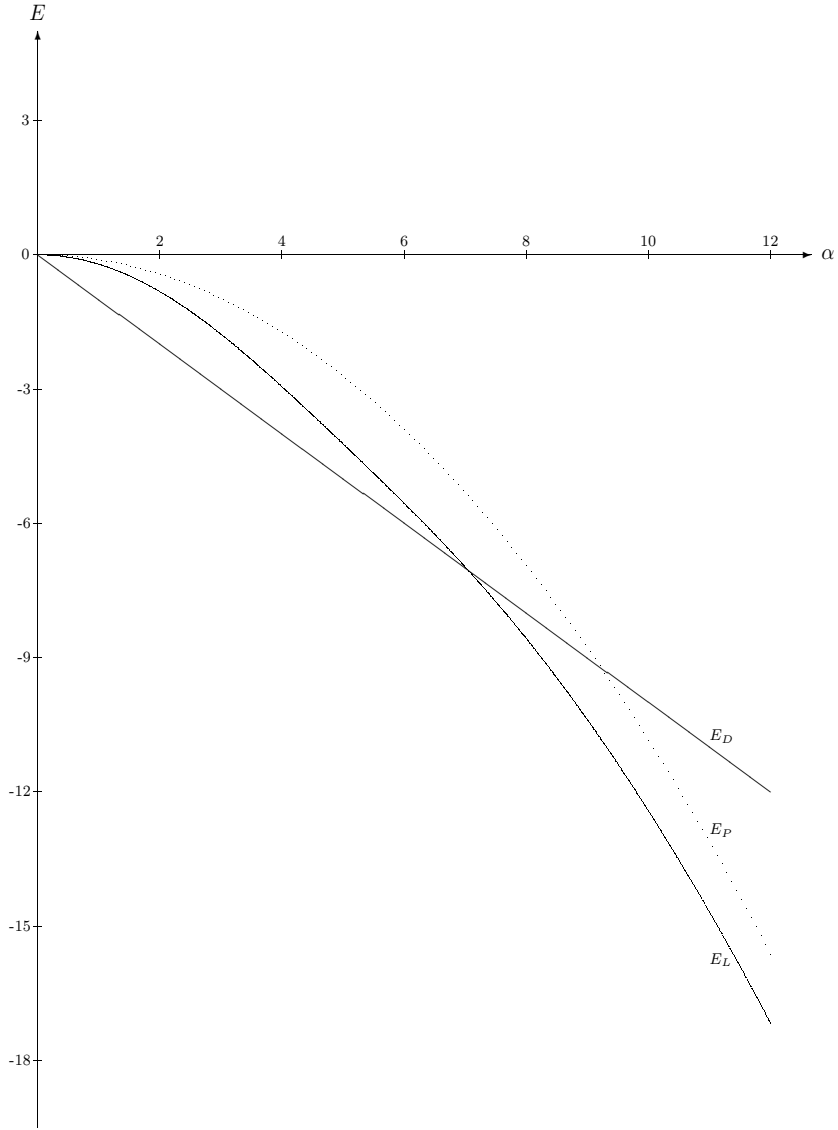}
\caption{Energies of polaron for Pekar's ($E_P$), delocalized ($E_D$)
and localized ($E_L$) states.}
\end{figure}

\section{New ansatz for the polaron state}

Fig.1 shows that generally speaking there
are no obvious restrictions on
the coupling constant $\alpha$ for both continuous functions
$E_{D}(\alpha)$ and $E_{L}(\alpha)$ corresponding to two
qualitatively different states.
The only suspicious point nearby $\alpha \simeq 7$
is defined by the intersection of these functions where
the states are degenerated. The similar situation is known in
the zone theory of electronic states in crystals as the
intersection of resonant zones \cite{13}. We use the same
method in order to analyze the transformation between these
states by means of the following linear combination

\begin{equation}
\label{eq16}
\vert \Phi_{\vec P}({\bf r})\rangle = C_{1} \vert \Psi^{(D)}_{\vec P}({\bf r})
 + C_{2} \vert \Psi^{(L)}_{\vec P}({\bf r})\rangle
\end{equation}
with only linear variational parameters $C_{1,2}$ and fixed
values of internal parameters for the basic states.

In a series of previous variational calculations of the polaron
binding energy for the intermediate coupling constant various
rather complicated trial functions for
the localized or delocalized state were used separately
(see, for example, the review \cite{14}) but the simple
state vector (\ref{eq16}) can't be reduced to any of them.

The subsequent calculations are rather bulky but standard.
By the definition the
both basic vectors are eigenstates for the operator of total
momentum and the ansatz (\ref{eq16}) should be used only for the
functional corresponding to the Schr\"odinger equation (\ref{eq3}).
The condition of vanishing the determinant obtained after variation
of this functional on the coefficients
$C_{1,2}$ leads to the following formula for two eigenvalues:

\begin{eqnarray}
\label{eq17}
\displaystyle{
E_{\pm}(\alpha)=\frac{1}{2(1-S^2)}\left(H_{DD}+H_{LL}-2SH_{LD}\pm E_R\right)}
\nonumber\\
E_R=\sqrt{(H_{DD}+H_{LL}-2SH_{LD})^2-4(1-S^2)(H_{DD}H_{LL}-H_{LD}H_{LD})}.
\end{eqnarray}

Here $H_{DL}, H_{DD}$ and $H_{LL}$ are the matrix elements
of the Hamiltonian
(\ref{eq1}) between the corresponding basic states and
$S$ is the ovelapping integral between this states being
nonorthogonal in general case. Omitting the calculations in detail
we write out the final analytical results for these values
when ${\bf P} = 0$.

The most simple value is

\begin{equation}
\label{eq18}
H_{DD} = - \alpha.
\end{equation}

The energy of localized state is defined by more complicated
formula

\begin{eqnarray}
\label{eq19}
\displaystyle{
H_{LL} = \frac{1}{N^2}\int d{\bf R}\int d{\bf r}\,\phi_0({\bf r}+{\bf R})
\exp(\sum_{\vec k}u^2_{\vec k}e^{-i\vec k\vec R})}
\nonumber \\
\displaystyle{
\Bigl\{-\frac{1}{2}\Delta+\sum_{\vec k}u^2_{\vec k}e^{-i\vec k\vec R}+
2^{3/4}\left(\frac{\pi\alpha}{\Omega}\right)^{1/2}\sum_{\vec k}
u^{\p+}_{\vec k}(e^{i\vec k\vec r}+e^{-i\vec k(\vec r +\vec R)})\Bigr\}
\phi_0({\bf r})}\\
\end{eqnarray}
with
$$
N^2=\int d{\bf R}\int d{\bf r}\, \phi_0({\bf r})\phi_0({\bf r}+{\bf R})
\exp(\sum_{\vec k}u^2_{\vec k}e^{-i\vec k\vec R}).
$$
At last
\begin{equation}
\label{eq20}
H_{LD}=\frac{e^{-\alpha/4}}{N}\int d{\bf r}\,\phi_0({\bf r})
\Bigl\{-\alpha +\frac{1}{2}\bigl(\sum_{\vec k}{\bf k}u_{\vec k}
v_{\vec k}\bigr)^2\Bigr\}
\exp\bigl(\sum_{\vec k}u_{\vec k}v_{\vec k}e^{i\vec k\vec r}\bigr)
\end{equation}
and
\begin{equation}
\label{eq21}
S=\frac{e^{-\alpha/4}}{N}\int d{\bf r}\,\phi_0({\bf r})\exp\bigl(\sum_{\vec k}
u_{\vec k}v_{\vec k}e^{i\vec k\vec r}\bigr).
\end{equation}




Formulae (\ref{eq17}) - (\ref{eq21}) permitted us to calculate the
functions $E_{\pm}(\alpha)$ in the entire range of the coupling
constant and Figure 2 shows the results. The well known estimation
for the polaron energy $E_{F}(\alpha)$ found by Feynman \cite{4}
is drawn for comparison. One can see that the functions
$E_{-}$ and $E_{F}$ are very similar continuous functions.
Certainly, $E_-$ becomes lower than $E_F$ for rather large $\alpha$
because of more accurate main asymptotical coefficient. The value
$E_-$  doesn`t depend essentially on the detailed form of the electron
function $\phi_0({\bf r})$ as one can see from the curve  $E_G$ corresponding
to the Gaussian approximation for this function.

\begin{figure}[ht]
\includegraphics[width=11cm,height=15cm]{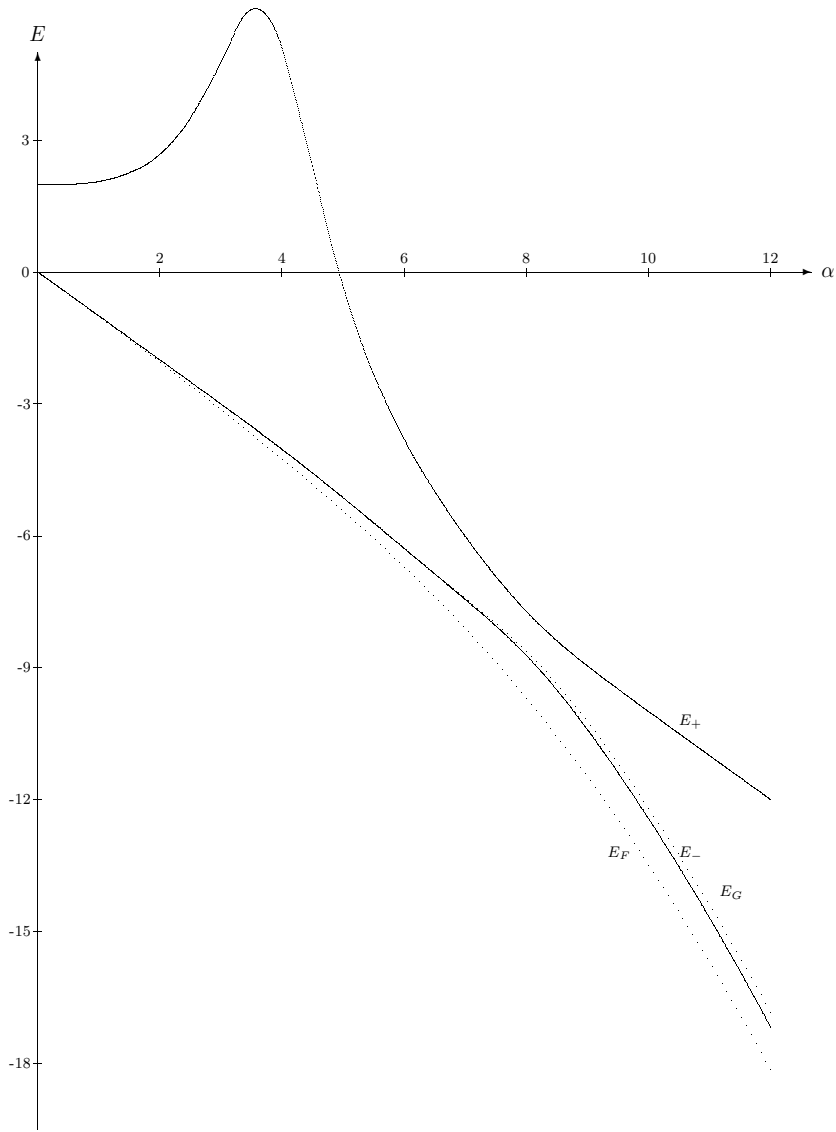}
\caption{Polaron binding energies calculated by Feynman ($E_F$)
and on the basis of our ansatz ($E_{\pm}$);
$E_G$ is the analog of $E_-$ but with the Gaussian
electron wave function.}
\end{figure}

The noticeable difference between our result and Feynman approximation in
the intermediate range is defined by more accurate constant term
in the strong-coupling series and it can be improved by means of the
second order approximation \cite{15}.
Figure 3 confirms it on the example of one-dimensional polaron with
Hamiltonian
\begin{equation}
\hat H=-\frac{1}{2}\frac{d^2}{dx^2}+\sum_k a^+_ka^{\p+}_k+
2^{1/4}\left(\frac{\alpha}{L}\right)^{1/2}\sum_k\left(e^{ikx}a^{\p+}_k+
e^{-ikx}a^+_k\right)
\end{equation}
where the role of the constant term is essentially less and we used
the exact solution of the nonlinear Schr\"odinger equation for
localized state.

\begin{figure}[ht]
\includegraphics[width=9cm,height=15cm]{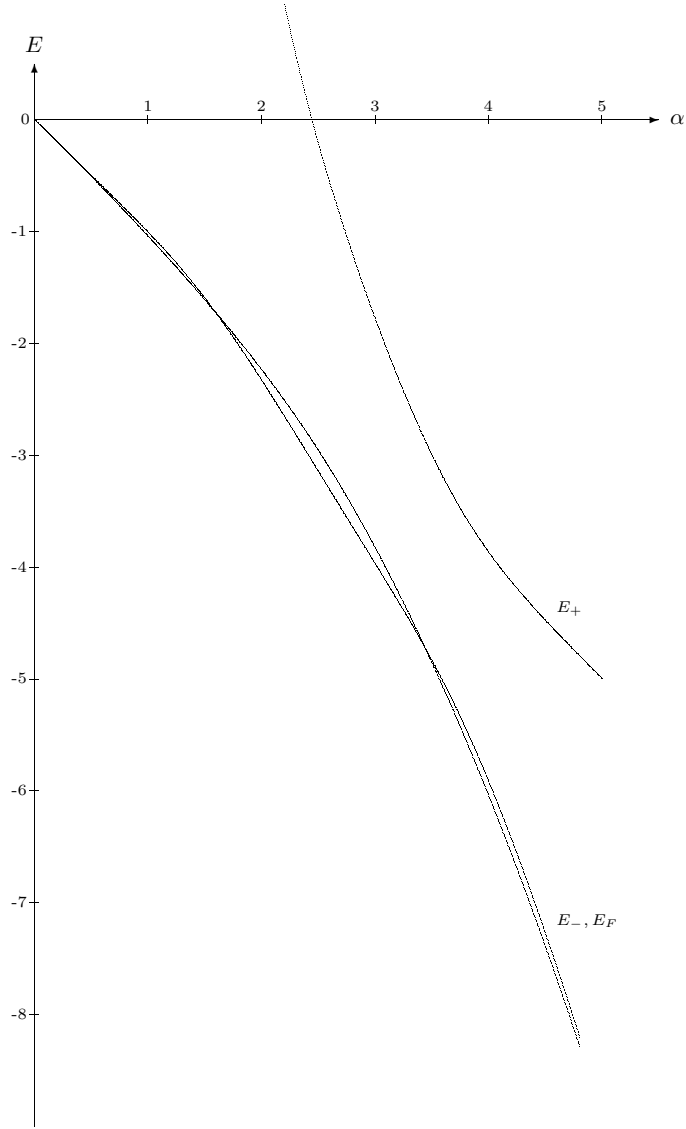}
\caption{The same as on Fig.2 but for the one-dimensional polaron.}
\end{figure}



\section{The model with divergencies in perturbation theory}

The polaron problem is the unique model because all diagrams of the
quantum perturbation theory for it are finite without renormalisation
procedure \cite{16}. Therefore it is of interest to apply our conception
of localized states to more general system. As an example, we consider
the following Hamiltonian
\begin{equation}
\label{eq22}
\hat H = -\frac{1}{2}\Delta+\sum_{\vec k}\omega_k a^+_{\vec k}a^{\p+}_{\vec k}
+\frac{g}{\sqrt{\Omega}}\sum_{\vec k}\frac{1}{\sqrt{2\omega_k}}
\left(e^{i\vec k\vec r}a^{\p+}_{\vec k}+e^{-i\vec k\vec r}a^+_{\vec k}\right),
\quad \omega_k=k=\vert{\bf k}\vert.
\end{equation}

It is easy to check that the particle self-energy diagram of the second
order is infinite in this model because of ultraviolet divergence of the
integral over {\bf k}. As a consequence, the delocalized state analogous
to (\ref{eq4}) does not exist here.

Let us now consider the localized states analogous to (\ref{eq13})
\begin{equation}
\label{eq23}
\vert\Psi({\bf r},{\bf R})\rangle=
\phi({\bf r}-{\bf R})\exp\left(\sum_{\vec k}\Bigl(u^{\p+}_{\vec k}e^{-i\vec k\vec R}
a^+_{\vec k}-\frac{1}{2}u^2_{\vec k}\Bigr)\right)\vert 0\rangle
\end{equation}
with
$$
u_{\vec k}=-\frac{g}{\sqrt{2\Omega\omega^3_k}}\int d{\bf r}\phi^2({\bf r})
e^{-i\vec k\vec r}
$$
and the wave function $\phi({\bf r})$ satisfies the nonlinear equation of the
same form (\ref{eq11}) as for polaron.

The total energy of this states
\begin{equation}
\label{eq24}
E_0=\int d{\bf r}\,\phi({\bf r})(-\frac{1}{2}\Delta)\phi({\bf r})-
\sum_{\vec k}\omega_k u^2_{\vec k}
\end{equation}
doesn`t contain the ultraviolet divergence due to the particle formfactor
and proves to be the finite value for any coupling constant $g$.

But there is another problem here connected with infrared divergence.
The matter is that the state vectors (\ref{eq23}) are not normalized
because of the term $\sum_{\vec k}u^2_{\vec k}$ which is proportional
to the total number of phonons and diverges over $k\to 0$.
The situation changes radically when we reconstruct the translational
symmetry of the system by analogy with (\ref{eq13})
\begin{equation}
\label{eq25}
\vert\Psi^{(L)}_{\vec P}({\bf r})\rangle=\frac{1}{N_{\vec P}\sqrt{\Omega}}
\int d{\bf R}\,\phi({\bf r}-{\bf R})\exp\left(i{\bf PR}+
\sum_{\vec k}\Bigl(u^{\p+}_{\vec k}e^{-i\vec k\vec R}-\frac{1}{2}u^2_{\vec k}
\Bigr)\right)\vert 0\rangle.
\end{equation}
The integral of normalization in this case (for simplicity ${\bf P}=0$)
is following
\begin{equation}
\label{eq26}
N^2_{\vec 0}=\int d{\bf R}\int d{\bf r}\,\phi({\bf r})\phi({\bf r}+{\bf R})
\exp\left(\sum_{\vec k}u^2_{\vec k}(e^{-i\vec k\vec R}-1)\right)
\end{equation}
and includes only the converging integrals.

Figure 4 shows the binding energy of the ``dressed'' particle

$$
E_L(g)=\int d{\bf r}\langle\Psi^{(L)}_{\vec 0}({\bf r})\vert\hat H
\vert\Psi^{(L)}_
{\vec 0}({\bf r})\vert,
$$

which is well defined and finite in the whole range of the coupling constant.
At the same time this value can`t be calculated on the basis of usual
perturbation theory because of the same reason as for the bound state in any
potential.

\begin{figure}[ht]
\includegraphics[width=15cm,height=15cm]{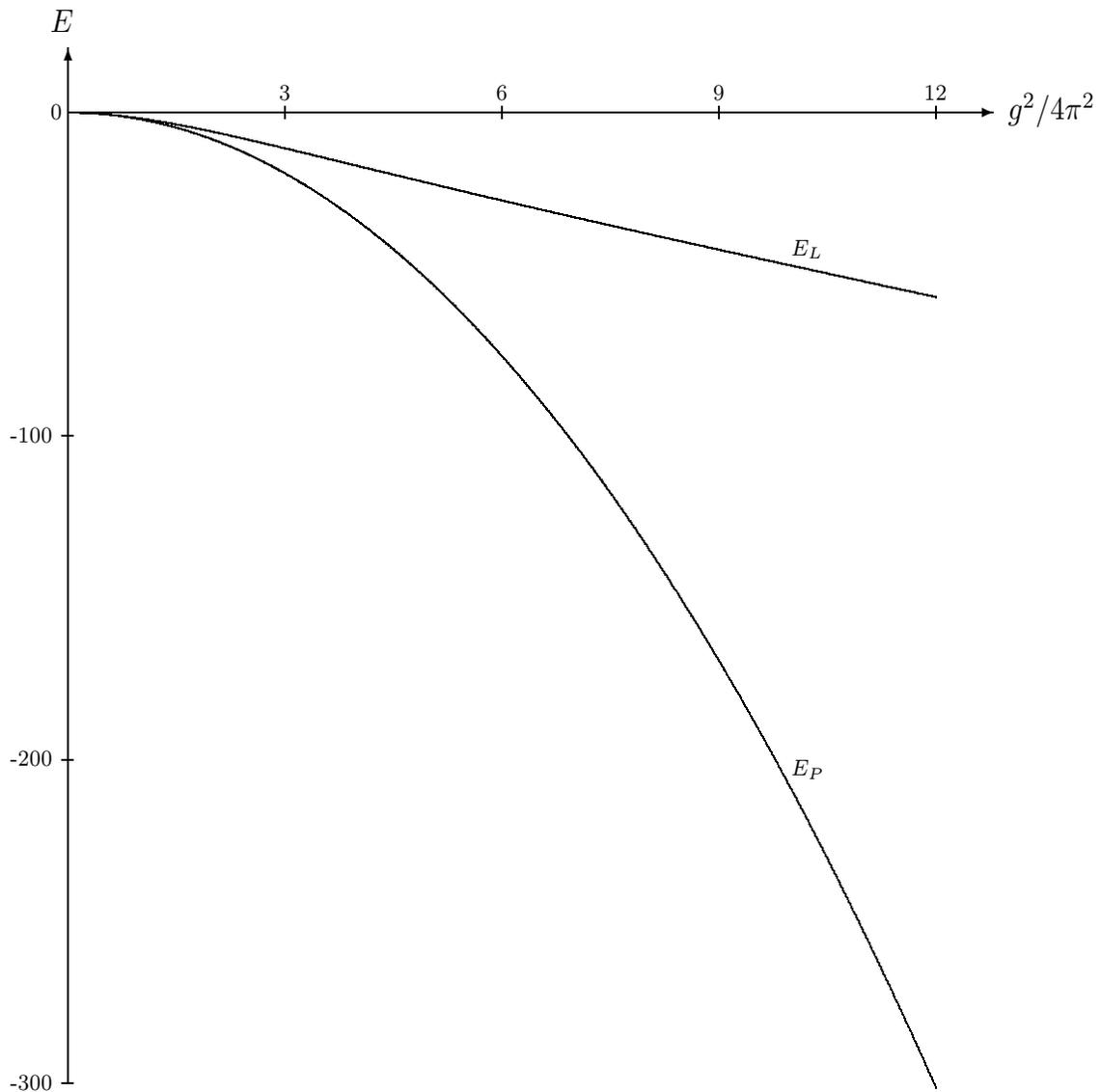}
\caption{Self-energy $E_L$ of the "dressed" particle for the model
with the Hamiltonian (21)
($E_P$ is the energy without
taking into account the translational symmetry).}
\end{figure}

\section{Conclusions}

The main purpose of this article is to make clear the question
of ``phase transition'' which appears for a number variational
approach used for description of the polaron in the Schr\"odinger
picture of the problem. Actually this question includes two
sides

(1) Do qualitatively different polaronic states really exist?
(2) Does the jump-like transition between these states take
place for some value of the electron-phonon coupling constant?

The new variational ansatz for the Schr\"odinger was found in our
paper and it was shown that the energy of
the polaron ground state
$E_{0}(\alpha)$ is continuous function of the coupling constant
in accordance with the analysis on the basis of the Feynman
path integrals. But at the same time there are two qualitatively
different states of the polaron without excitation of real phonons.
The pseudo-intersection between these states describes in detail
the transition between the delocalized state of the electron in
crystal for small $\alpha$ to the localized one in the strong
coupling limit and can lead to some physical effects nearby the
point of quasi-degeneracy.

One of the advantages of the considered operator calculation of
the value  $E_{0}(\alpha)$  is defined by the fact that we
use only the eigenfunctions of the total momentum operator
and it permits us to calculate the effective mass of the polaron
by consistent way. We'll consider the corresponding rather bulky
calculation separately. Besides, the operator approach allows to find
the exact asymptotic serieses in the both limit cases on the coupling
constant and calculate by regular way the subsequent corrections in
the intermediate range.

And the important result of the paper is defined by the generalization
of the conception of localized states in application to the system with
the ultraviolet and infrared divergencies of the perturbation theory.
We beleive that the considered non-perturbative description of the
``dressed'' particle is of great interest for quantum field theory.

\section{Acknowledgments}

We are grateful to the Belarussian Foundation for Fundamental
Research (Grants No Ф97/244 and No Ф97/099) and Deutscher Akademischer
Austauschdienst for the support this work.

\end{document}